\documentclass{elsart}

\newcommand{\RA}[3]{{#1}^{{\rm h}}{#2}^{{\rm m}}{#3}^{{\rm s}}}
\newcommand{\Dec}[3]{{#1}^{\circ}{#2}'{#3}''}
\newcommand{\gsim}{\mbox{\hspace{.2em}\raisebox{.5ex}{$>$}\hspace{-.8em}\raisebox{-.5ex}{$\sim$}\hspace{.2em}}}
\newcommand{\lsim}{\mbox{\hspace{.2em}\raisebox{.5ex}{$<$}\hspace{-.8em}\raisebox{-.5ex}{$\sim$}\hspace{.2em}}}

\usepackage{amssymb}
\usepackage{epsfig}
\usepackage{natbib}

\begin{document}
\begin{frontmatter}


\title{The Heart-shaped Supernova Remnant 3C~391 viewed in Multi-bands }

\author{Yang Su, Yang Chen}

\address{Department of Astronomy, Nanjing University,
Nanjing 210093, P.R.\ China}

\begin{abstract}
Using {\sl Chandra} X-ray, {\sl Spitzer} mid-IR, and 1.5~GHz radio data,
we examine the spatial structure of SNR~3C~391.
The X-ray surface brightness is generally anti-correlative with
the IR and radio brightness.
The multiband data clearly exhibit a heart-shaped morphology
and show the multi-shell structure of the remnant.
A thin brace-like shell on the south detected at 24~$\mu$m is projected outside
the radio border and confines the southern faint X-ray emission.
The leading 24~$\mu$m knot on the SE boundary appears to be partly
surrounded by soft X-ray emitting gas.
The mid-IR emission is dominated by the contribution of the shocked
dust grains, which may have been partly destroyed by sputtering.

\end{abstract}

\end{frontmatter}

\section{Introduction}
Galactic molecular clouds, containing about half the mass of the interstellar 
medium (ISM), have a complex structure, and stars form out of the densest 
regions therein. Massive stars have often not moved far from their
birthsites by the time they explode as a core-collapse supernova, and
therefore the interaction of supernova remnants (SNRs)
with molecular clouds can plausibly occur.
The number of SNRs with convincing evidence for interaction with ambient
molecular clouds is about 20,
and a large fraction of them are mixed-morphology or thermal composite
SNRs, with interior-filled thermal X-ray emission surrounded by a radio shell
(Yusef-Zadeh et al.,\ 2003).

When SNRs interact with inhomogeneous molecular gas with a large
density variation, the
morphology and structure is not expected to be regular or symmetric.
Features such as gas clumps, incomplete arcs, blowout shells,
thin or thick filaments, etc., can typically be observed in radio, 
infrared (IR), optical, or X-rays.
SNR 3C~391 is one of the prototype mixed-morphology or thermal composite SNRs,
confirmed to be interacting with molecular clouds to the northwest
(Rho \& Peter, 1998; Wilner et al., 1998; Reach \& Rho, 1999; 
Chen \& Slane, 2001).
Due to its remarkable radio shell elongated  from
northwest (NW) to southeast (SE), it is suggested to have broken
out of a dense region into an adjacent region
of low density (Reynolds \& Moffett, 1993).
The bright clumpy thermal X-ray emission arises from the remnant interior,
and the rim is X-ray faint (Chen \& Slane, 2001;
Chen et al., 2004, hereafter CSSW04).
The {\sl Spitzer Space Telescope (SST)} (Werner et al., 2004)
{\sl Infrared Array Camera (IRAC)}
near-IR observations show a bright incomplete shell
correlated well with the NW radio structure which is tangent
to a giant molecular cloud (Reach et al.,\ 2005; Lee, 2005).
These structure patterns at different wavelengths show complex physical
conditions in SNR 3C~391, and reflect inhomogeneities in the ISM.

Here we will show as overall view of SNR 3C~391 combining the
radio (1.5 GHz), mid-IR ({\sl SST} 24 \& 70~$\mu$m), and X-ray ({\sl Chandra})
observations,
and examine the spatial structures in different wavelengths.

\section{\label{mul}Multi-band observation of 3C~391}
\subsection{The Radio and X-ray Morphology}
In the radio images (Reynolds \& Moffett, 1993; Moffett \& Reynolds, 1994),
3C~391 has a bright partial shell along the north and NW boundary.
The NW part of the remnant appears to be a broken bubble,
with the brightest radio bar along the western border and a
faint blowout extending to the SE.
Two 1720 MHz OH maser spots are detected along the radio shell
(Frail et al., 1996).
Reynolds \& Moffett (1993) suggest that the remnant may have resulted from
a supernova explosion in the interior of the molecular cloud and
is expanding into a region of greatly varying density.

The {\sl Einstein} (Wang \& Seward, 1984) and {\sl ROSAT} (Rho \& Peter, 1996) 
X-ray observations show a strong soft X-ray emission arising from the 
southeast region of the remnant.
Both the {\sl ASCA} and {\sl Chandra} observations show a SE-NW elongated
broad-band X-ray morphology, similar to the radio image
(Chen \& Slane, 2001; CSSW04).
The X-ray emission peaks southeast of the geometric center
of the radio pattern.
The X-ray emission from the interior to the NW bubble suffers higher
absorption than that from the SE blowout part.
The {\sl Chandra} observation reveals rich small-scale structures,
such as knots and arcs, as well as faint, diffuse gas that
appears to expand out of the southwestern radio boundary (CSSW04).
In the southwest (SW), the narrow band analysis of the S and Si X-ray lines
unveils a very faint jet-like protrusion projected outside
the radio border (Su \& Chen, 2005).

To better show the entire spatial X-ray emitting remnant, we
rebin the cleaned level 2 {\sl Chandra} data (ObsID 2786; CSSW04)
using an adaptive mesh to include at least 9 counts
(0.3-7 keV) per bin. The contours of the rebinned image are plotted
in Fig.~1a. In this figure, a broad band X-ray protrusion
(corresponding to the narrow band one) is seen in the SW.

\subsection{Mid-IR images}
The mid-IR 24~$\mu$m and 70~$\mu$m observations used here were
carried out as part of the 24 and 70 Micron Survey of the Inner
Galactic Disk Program (PID: 20597, PI: Sean Carey) with the 
Multiband Imaging Photometer (MIPS) (Rieke et al., 2004).
The data are obtained from {\sl Spitzer} archive.
The raw 24~$\mu$m image of 3C~391 from the Post Basic Calibrated
Data is shown in Fig.~1b, which displays the remnant extent which is similar
to that seen in radio and X-rays. 
We sum 24296 pixels ( $\sim3.7$ arcmin radius) to get the 
total surface brightness and select faint regions outside the SNR as
the background. Subtracting the mean 
surrounding IR background  $\sim60\pm3$~MJy/sr, we obtain a mean 
surface brightness of 3C~391 $\sim11.5\pm3.0$~MJy/sr or a 
flux $\sim37.9\pm9.9$~Jy.
After a manual subtraction of
nearby emission features, the remaining 24~$\mu$m emission is
superposed on the X-ray contours in Fig.~1a. The manually
subtracted 24~$\mu$m and 70~$\mu$m emission maps are overlaid with
the 1.5~GHz radio contours in Fig.~2. Tricolor image with 1.5~GHz
radio emission in red, 24~$\mu$m IR emission in green, and
0.3--7.0~keV X-rays in blue is shown in Fig.~3.
In the 70~$\mu$m mid-IR
image there are some streaks mixed with the IR emission and it is
very difficult to determine the background intensity or the pure
mid-IR intensity associated with SNR. We use the GeRT package 
(http://ssc.spitzer.caltech.edu/mips/gert) and the software tool MOPEX 
(http://ssc.spitzer.caltech.edu/postbcd/download-mopex.html) to creat a 
new mosaic of 70~$\mu$m. After running MOPEX on the filtered Basic 
Calibrated Data (BCD) images produced by the 
data processing pipeline, some arc-like structures 
of 3C~391 are visible in 70~$\mu$m (Fig.~2b).

The two mid-IR images (Fig.~2) display several thick arc-like
structures along the SW, NW, and northeast (NE) boundarier.
A very faint 24~$\mu$m
brace-symbol-like thin arc structure in the south is projected
outside the radio border and goes from west to south along the
edge of the faint X-ray emission.

In the 24~$\mu$m emission map (Fig.~1a, 1b), there are two bright knots at
($\RA{18}{49}{39}.9,\Dec{-00}{56}{40}$) and ($\RA{18}{49}{39}.4,
\Dec{-00}{58}{36}$) on the east and southeast border, respectively.
They seem to correspond to X-ray knots (Fig.~3). The east knot
is the 24~$\mu$m counterpart of IR source IRAS 18470$-$0100 (CSSW04),
with improved positional resolution.
It can also be seen that a line of point-like IR sourses are evenly
spaced extending from the remnant center to the
bright SE knot. These five sources have {\sl 2MASS} 
near-IR counterparts with accurate positions:
$(\RA{18}{49}{27}.17$, $\Dec{-00}{56}{44}.1)$, 
$(\RA{18}{49}{29}.34$, $\Dec{-00}{57}{29}.9)$, 
$(\RA{18}{49}{31}.43$, $\Dec{-00}{58}{02}.5)$,
$(\RA{18}{49}{35}.15$, $\Dec{-00}{58}{26}.1)$, and
$(\RA{18}{49}{39}.50$, $\Dec{-00}{58}{36}.0)$. 
The IR source at the geometric center has an X-ray counterpart
(CXJ~184927.0-005640; CSSW04) given $\sim4''$ spatial
resolution of {\sl 2MASS}.
The nature of these sources is interesting but unknown yet.
If the central one suffers an interstellar absorption similar
to that of 3C~391 ($N_H\sim3\times10^{22}$~cm$^{-2}$),
assuming its X-ray spectrum as a thermal plasma with a temperature
of 0.6 keV (for soft case) and 2 keV (for hard case), respectively,
typical for normal early-type stars, we estimate
the $0.3-8$ keV luminosity of the X-ray source as
$\sim4.3\times10^{32}$~ergs~s$^{-1}$ (for soft case) and 
$\sim3.3\times10^{33}$~ergs~s$^{-1}$ (for hard case) based on 
the X-ray count rates (Table~1 in CSSW04).
It indicates that, on this assumption, the central source would most
probably represent an individual early-type star or even a 
massive colliding stellar wind binary.

\begin{figure*}
\centerline{\epsfig{file=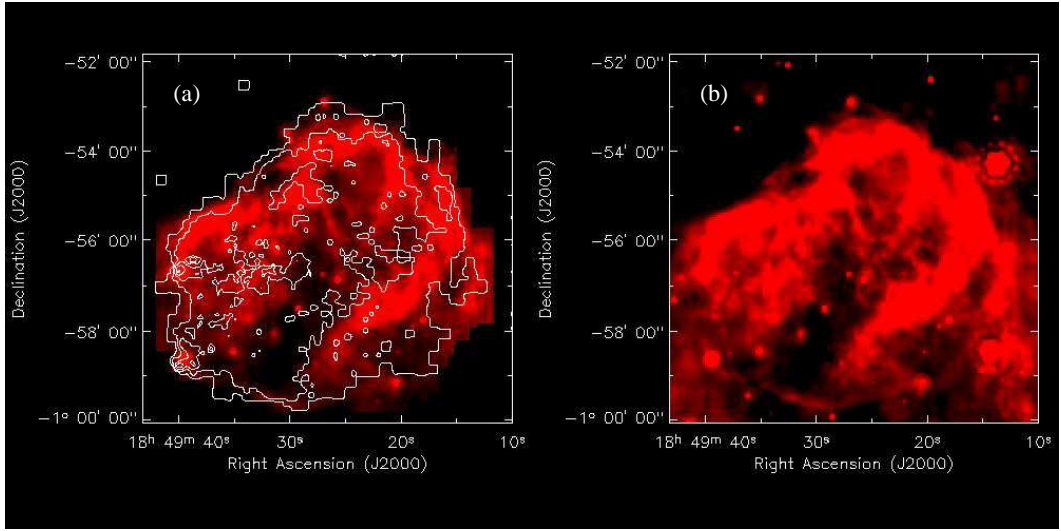,width=14cm}}
\caption{
(a) Superposition of {\sl Chandra} X-ray (0.3-7.0 keV) contours of 3C~391
on the MIPS 24~$\mu$m image.
The X-ray contours are between the 5$\%$ maximum and the 90$\%$ maximum
with a square-root scale.
The IR intensity range is from 62~MJy sr$^{-1}$ to 80~MJy sr$^{-1}$,
with a linear transfer function and the point-like IR sources away from the SNR
have been removed.
(b) The {\sl SST} MIPS 24~$\mu$m raw image of SNR 3C~391.
The IR intensity range is from 58~MJy sr$^{-1}$ to 92~MJy sr$^{-1}$, with
a linear transfer function but the background and the point-like sources
have not been removed.}
\end{figure*}

\begin{figure*}
\centerline{\epsfig{file=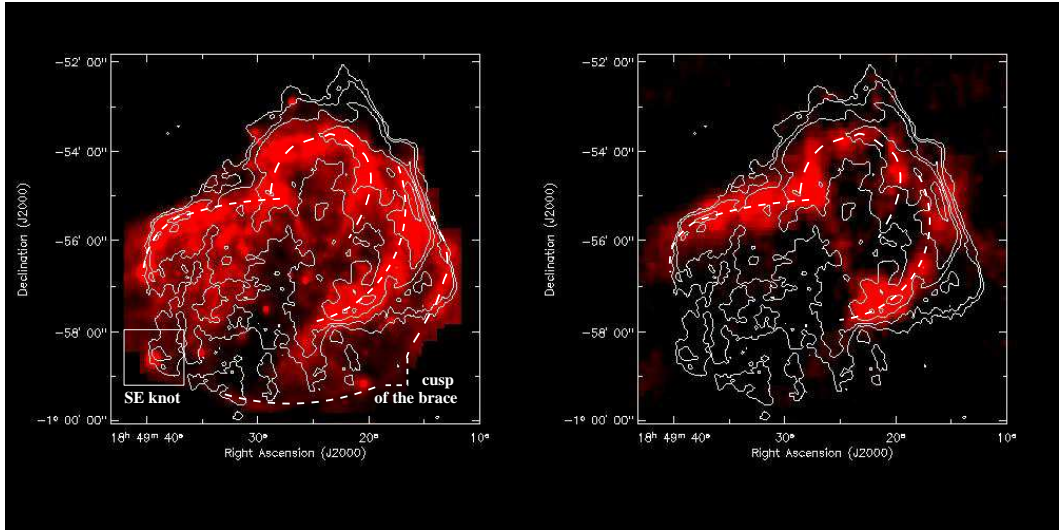,width=14cm}}
\caption{
(a) The {\sl SST} MIPS 24~$\mu$m image of SNR 3C~391,
overlaid with the 1.5 GHz radio contours.
The IR intensity range is from 62~MJy sr$^{-1}$ to 80~MJy sr$^{-1}$,
with a linear transfer function and the point-like IR sources away the SNR
are removed. The overlaid 1.5 GHz radio contours are at 1.5, 3.19, 8.25,
16.69, 43.68, and $62.24 \times 10^{-3}$ Jy/beam
(from Moffett \& Reynolds 1994).
(b) The MIPS 70~$\mu$m image of SNR 3C~391 overlaid with
the same radio contours as used in (a). 
The radio character of
the image is same as in (a). The dashed lines indicate pronounced
IR arc/shell-like structures.}
\end{figure*}

\begin{figure*}
\centerline{\epsfig{file=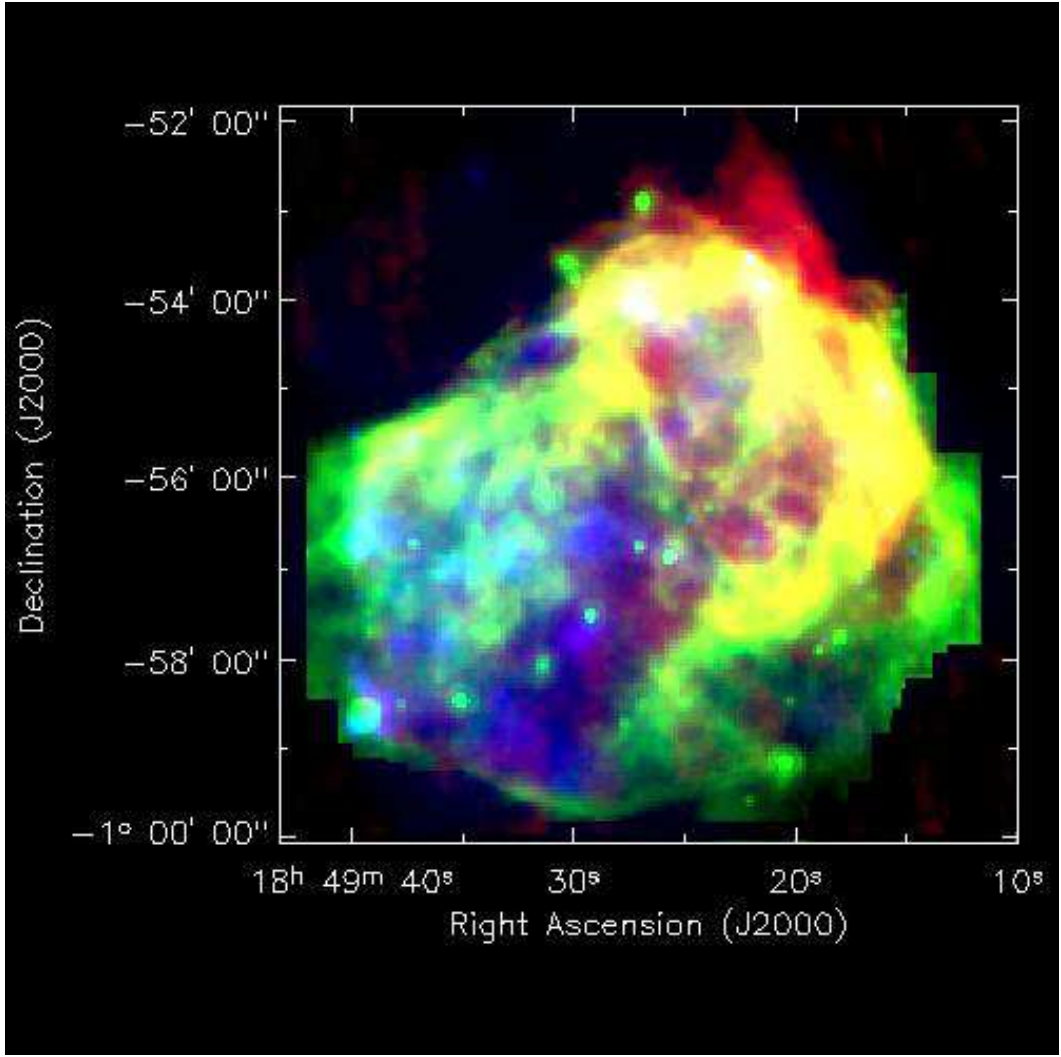,width=14cm}}
\caption{
Tricolor image of SNR 3C~391. The 1.5 GHz radio is coded in red,
the {\it SST} 24~$\mu$m IR in green, and the {\sl Chandra} X-ray in blue.
The intensity map is square-root scaled in each energy band.
}
\end{figure*}

\begin{figure*}
\centerline{\epsfig{file=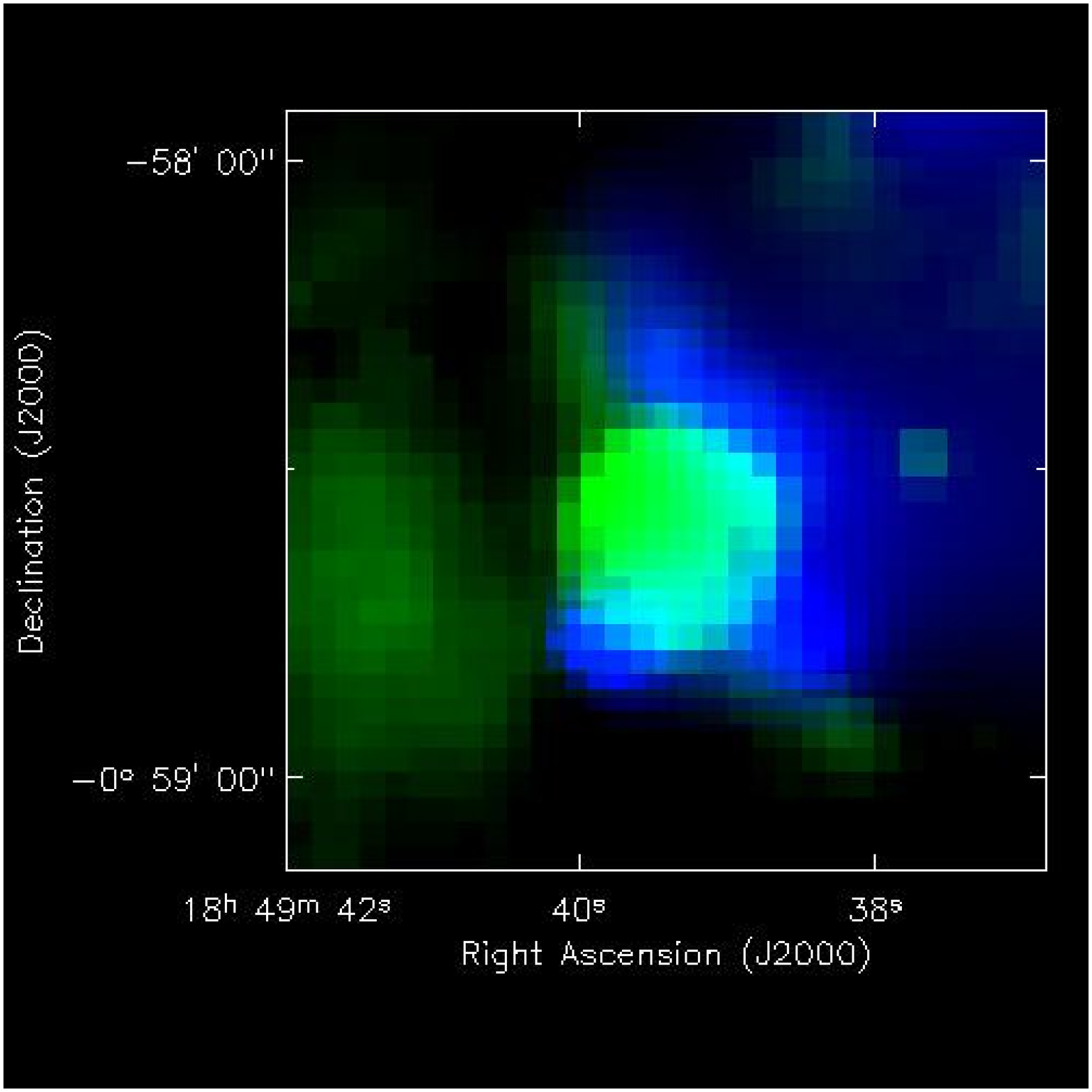,width=7cm}} \caption{ The
southeastern boundary region. The 24~$\mu$m IR emission is coded
in green, and the {\sl Chandra} soft (0.3-2.6 keV) X-ray emission in blue. }
\end{figure*}

\section{Discussion}
\subsection{Spatial comparison among X-ray, mid-IR, and radio emissions}
Using the available radio, IR, and X-ray data, we can investigate the 
structure of SNR 3C~391.
Seen in the tricolor image (Fig.~3), the X-ray surface brightness
is generally anti-correlative with the IR and radio brightness,
or anti-correlative with the environmental density
because of intervening extinction.
In the NW half of the remnant, two distinct thick
arc-like structures in both 24 and 70~$\mu$m coincide grossly with
the radio shell (Fig.~2 and Fig.~3).
The westernmost bar-like radio peak and the ``broad molecular line"
(BML) region are located along the west-southwest
IR shell, which represents a dense shocked region.
The two 1720 MHz OH maser spots (Frail et al., 1996) are harbored
in the NE and SW thick radio/mid-IR shells.
On the NE boundary, a thick mid-IR arc is coincident with the radio
shell, too.
These arcs/shells are also discernible in the {\sl SST} 5.8~$\mu$m
near-IR images (Reach et al., 2005; Lee, 2005).

Apart from the above general comparison, we note two features of
particular interest.

1. The knot on the southeastern border $(\RA{18}{49}{39}.4$,
$\Dec{-00}{58}{36})$ (the SE knot in Fig.~2a) appears to be 
partly surrounded by soft
X-rays (Fig.~4). Two possible mechanisms may explain this small-scale
phenomenon.
First, after the blast strikes a small dense cloud, a transmitted
shock heats the cloud to emit 24$\mu$m emission, while a
bow-like reflected shock stands around the cloud (McKee \& Cowie,
1975). Second, after the cloud is shocked by the
blast wave, materials are evaporated from the cloud and form a
soft X-ray emitting gas layer (Cowie \& McKee, 1977).


2. The 24~$\mu$m brace-like thin shell (Fig.~2a) appears to be projected
outside the long
southern radio border of the remnant (Fig.~2a and Fig.~3). It has
not been seen before at any wavelengths. However, it
confines very well the faint X-ray emitting emission that fades out
from the remnant interior and seems to extend beyond the SW radio
border (CSSW04) (Fig.~1a and Fig.~3).
Moreover, the position of the cusp of the ``brace'' (Fig.~2a) is coincident 
with a jet-like protrusion seen in the narrow band X-ray maps for S
and Si (Su \& Chen, 2005).
Such positional consistency between the 24~$\mu$m and
X-ray emission suggests that the thin brace-like IR shell is
associated with the remnant.
Thus, the 3C~391 SNR has a larger
volume than previously recognized and takes a heart-like shape with
combination of the X-ray, IR, and radio emission (Fig.~2a and Fig.~3).
The multi-shells, especially the newly detected southern thin shell,
indicate that the molecular clouds around this remnant are
distributed more complexly than previously assumed.
The NW clouds have a non-uniform structure and 
there is a low density region in the south, into which a part
of the supernova blast wave propagates.

\subsection{The 24~$\mu$m mid-IR emission of the multi-shells}
The 24~$\mu$m emission more distinctly demonstrates shell structures
of SNR~3C~391 than other IR wavelength emission.
Similar phenomena occur in some other SNRs (Stanimirovic et al., 2005;
Morris et al., 2006; Borkowski et al., 2006).
Hence the 24~$\mu$m band seems to be an effective window for
SNR observation.
The near-IR emission from various portions of the 3C~391 SNR has been suggested
to be dominated by atomic fine-structure lines and synchrotron emission 
(from electrons in the moderate-density molecular regions)
or molecular lines and ionic lines (from the high
density molecular gas) (Reach et al., 2005).
Here we suggest that the mid-IR emission of 3C~391 mainly arises from
dust grains.

The origin of the 24~$\mu$m mid-IR emission is not clear. We explored 
possible mechanisms of synchrotron, atomic lines, molecular lines, or
IR emission from shock-heated dust grains.
Let us compare the contribution
of each mechanism to the 24~$\mu$m emission of 3C~391.
(a) {\sl Synchrotron.}
Using the integrated flux at 330 MHz of 41.0~Jy and the spectral
index $\sim-0.49\pm$0.1 (Brogan et al. 2005),
we estimate the flux by
extrapolation to be less than 0.3 Jy (1\% total) at 24~$\mu$m.
Certainly, the mid-IR emission of the brace-like thin shell 
is not synchrotron, otherwise it should have been discernible at
70~$\mu$m.
(b) {\sl Ionic lines.}
The {\sl SST} MIPS band at 24~$\mu$m contains ionic lines
of [FeII] at 25.988~$\mu$m and [OIV] at 25.890~$\mu$m (Reach \& Rho, 2000).
In the BML region and the radio bar-like peak, 
the [FeII] plus [OIV] lines amount to no more than 40\% and 30\% of the
local MIPS IR 24~$\mu$m fluxes, respectively (Reach \& Rho, 2000).
We assume that similar fractions hold for the entire remnant
but note that this may overestimate the fraction because the two
regions are the brightest in the ionic lines.
(c) {\sl Molecular lines.}
Some rotational molecular lines such as
from OH and H$_2$O may contribute to the mid-IR emission,
but generally mid-IR from molecular lines is weak compared with the
strong continuum emission of dust grains (Gorti \& Hollenbach, 2004).
Also, the ISO observations of this remnant do not show
strong molecular lines at 24~$\mu$m (e.g., Reach \& Rho, 2000).
(d) {\sl Dust emission.}
Since the 24~$\mu$m emission of 3C~391 is not dominated by any of
the above three mechanisms, we suggest that it predominantly ($>60\%$)
originates from dust grains. 
The concentration of the 24~$\mu$m emission on the remnant shells
also favors that the dust in SNR~3C~391 is part of the swept-up,
shock-heated ISM. This physics of explanation is similar to that
for the mid-IR emission of SNR~N132D (Tappe et al., 2006).

The blackbody temperature of the dust grains in SNR~3C~391
was estimated as 140~K for 12-24~$\mu$m and 52~K for 24-100~$\mu$m
from the {\sl IRAS} observation (Arendt 1989).
The total mass $M_d$ of the hot dust associated with
the SNR at temperature $T_d$ is given by:
\begin{equation}
M_d = { d^2 F_{\nu}(\lambda) \over \kappa(\lambda) B_{\nu}(\lambda,\ T_d)}
\qquad ,
\end{equation}
where $F_{\nu}(\lambda)$ is the observed 24~$\mu$m flux ($\gsim23$~Jy),
$d\sim$8.0~kpc the distance to the remnant (Chen \& Slane 2001),
$\kappa_{\lambda}=2.5(\lambda/{\rm 450}\mu m)^{-2.0}$ cm$^2$ g$^{-1}$
the dust mass absorption coefficient (Draine \& Lee 1984),
and $B_{\nu}(\lambda,\ T_d)$ the
Planck function evaluated at the dust temperature.
Thus we get a lower limit on the dust mass $M_d\gsim 2\times10^{-4}\,M_{\odot}$
for $T_d\sim140$~K and $M_d\gsim 0.3\,M_{\odot}$ for $T_d\sim52$~K.
With the mass of the X-ray emitting gas of the remnant
$M_{g}\sim 100~\,M_{\odot}$ (CSSW04),
the dust-to-gas mass ratio is then $\gsim2\times10^{-6}$--$3\times10^{-3}$
 (for 60\% infrared flux accounted) and $\lsim3\times10^{-6}$--$5\times10^{-3}$
 (for 100\% infrared flux accounted).
The estimated ratio is smaller than the normal Galactic
value ($\sim0.01$; Spitzer, 1978),
implying that a fraction of the shocked dust grains have been
destroyed by sputtering. 
Given the lifetime of the dust grains destroyed by sputtering
for gas temperature $\gsim10^{6}$ K,
$\tau_{\rm sput}\approx 10^{6}a(\mu {\rm m})/n({\rm cm}^{-3})$~yr
(where $a$ is the grain radius; Dwek \& Werner 1981),
and adopting the gas density $n\sim$2 cm$^{-3}$ for the 3C~391 (CSSW04),
we could essentially expect that 
some large grains ($a\gsim0.01\mu$m) which were initially present
survive for the age of the remnant ($\sim4\times$ 10$^{3}$~yr) (CSSW04).

\section{Summary}
We have compared the morphology of SNR 3C~391 with the data of
{\sl Chandra} X-ray, {\sl SST} mid-IR, and 1.5~GHz radio observations.
The X-ray surface brightness is generally anti-correlative with
the IR and radio brightness.
The combination of the three band data clearly exhibits an multi-shell
structure and a heart-shaped entire morphology for the remnant.
Some detailed structures are revealed:\\
1.\ A thin brace-like shell detected at 24~$\mu$m, which is projected
outside the radio border confines the southern faint X-ray emission.
The cusp is coincident with sn X-ray protrusion.\\
2.\ The 24~$\mu$m knot on the SE boundary appears to be partly
surrounded by soft X-rays.\\
The mid-IR emission is probably dominated by dust grains, which are heated
by the shocked hot gas and may have been partly destroyed by sputtering.

{\bf Acknowledgements:}
We have benefited from the archived {\sl Spitzer} 24 and 70 Micron Survey
of the Inner Galactic Disk Program.
Two anonymous referees are thanked for useful comments that led to
various improvements in the presentation of this paper.
Y.C.\ acknowledges support from NSFC grants 10221001 and 10673003.

\end{document}